\documentclass[prl,aps,twocolumn,nofootinbib,floatfix]{revtex4}
\usepackage{epsf,amssymb,amsmath,mathptmx}
\begin{document}
\maketitle
\noindent
{\bf Goh et al. Reply:} We introduced in a recent Letter~\cite{goh}
the load distribution following a power law on scale-free (SF) networks.
In addition, it was conjectured that the load exponent $\delta$
is universal as long as the degree exponent $\gamma$ is in
$2 < \gamma \le 3$, based on real-world networks and
{\it in silico} models.
In the preceding Comment~\cite{barth}, Barth\'elemy argues that
$\delta$ is not universal, sensitive to the details of SF networks. 
In this reply, we notice that the discrepancy is mainly caused by 
different usages of definition of load in~\cite{goh} and~\cite{barth}. 
Following the definition used in~\cite{barth}, we agree with the result 
of~\cite{barth}, however, we find that the question of the universality 
of the load exponent is not settled yet. 

\begin{figure}[b]
\centerline{\epsfxsize=8.0cm \epsfbox{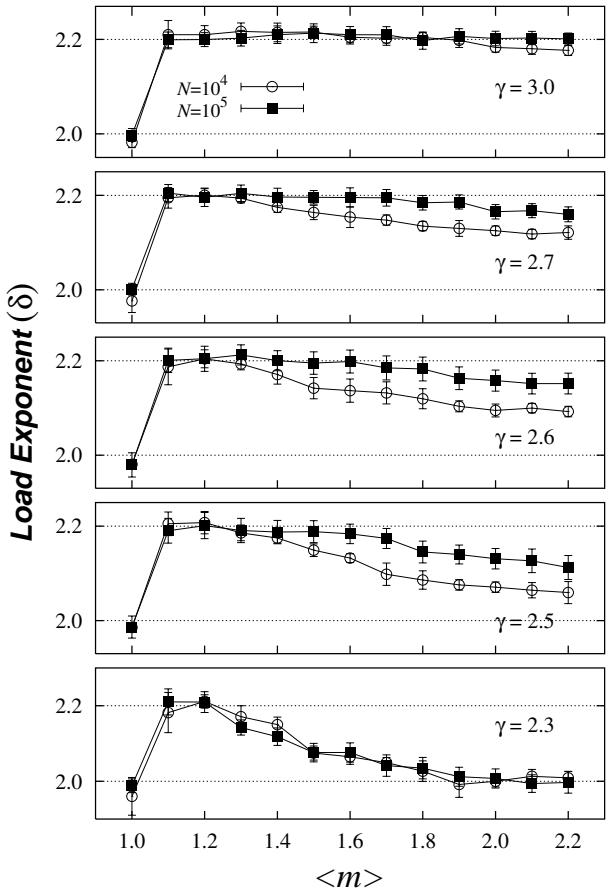}} 
\caption{The load exponent as a function of the mean number 
of edges $\langle m \rangle$ emanating from a new vertex 
for various degree exponents $\gamma$ in the BA model and 
different system sizes, $N=10^4$ ($\bigcirc$) and 
$N=10^5$ ($\blacksquare$).}
\end{figure}

In~\cite{goh}, the load $\ell_k$ of a vertex $k$ includes $N-1$
packets leaving and another $N-1$ packets arriving at the vertex,
where $N$ is the total number of vertices. However, those $2(N-1)$
packets are not included in~\cite{barth}. While the difference of
$2(N-1)$ can be neglected for vertices with large load $\ell$ in
the limit of $N\rightarrow \infty$, however, in finite-size
systems particularly those compatible with most real-world
networks comprising $N=10^3\sim 10^4$ vertices, this difference
could produce a different value of $\delta$. We perform extensive
numerical simulations on a larger scale $N=5\times 10^5$ than the
size $N=10^4$ previously used in [1] for the static model with
$\gamma \approx 2.5$, following the definition in~\cite{goh}, 
and find that indeed
$\delta$ turns out to be lower than $\delta \approx 2.2$ beyond 
error bar as argued in~\cite{barth}. 
This behavior also occurs in the model
introduced by Barab\'asi and Albert (BA) when the number of edges
emanating from a newly added vertex is $m \ge 2$ in finite-size
systems. However, we will show that the universal behavior of 
the load exponent is still likely as far as SF networks are sparse.

The load exponent for SF tree has been obtained
analytically to be $\delta=2.0$, independent of the degree
exponent $\gamma$~\cite{pnas,szabo}. We investigate how the
exponent value $\delta=2.0$ changes as the number of loops
increases. We modify the BA model in such a way 
that a new vertex attaches one or two edges to existing network 
with probability $1-p$ or $p$, respectively. The mean number of
edges emanating from a new vertex is then $\langle m \rangle=1+p$.
We investigate how the load distribution changes as $\langle m
\rangle$ varies. When $p=0$, the network is 
tree, and the load exponent is confirmed to be $\delta\approx
2.0$. We find that $\delta$ increases to $\delta\approx 2.2$ by
increasing $\langle m \rangle$ to $\langle m \rangle\approx 1.1$
at which the edges connecting different branches of the
tree structure form sparse loops in a nontrivial manner. 
The value $\delta\approx 2.2$ turns out to be robust, 
independent of the degree exponent $\gamma$ for $2 < \gamma < 3$. 
Such behavior persists as long as $\langle m \rangle$ is smaller than 
a $\gamma$-dependent critical value, $\langle m \rangle_c$, 
beyond which $\delta$ depends on $\gamma$ as observed in~\cite{barth}. 
Moreover, we find that the plateau region of $\delta\approx 2.2$ 
is extended as the system size $N$ increases as shown in Fig.~1. 
These data suggest that the universal behavior of $\delta$ may 
hold in some finite region of parameter space in the
thermodynamic limit, at least for the sparse BA model. 
Thus the possibility of the universal behavior of the load exponent 
is still an open question. Further details will be published 
elsewhere~\cite{dkim}.

This work is supported by the ABRL program of the KOSEF.\\

\noindent K.-I. Goh, C.-M. Ghim, B. Kahng, and D. Kim\\
\small{
\indent School of Physics and Center for Theoretical Physics,\\
\indent Seoul National University, Seoul 151-747, Korea.\\
\vskip -.1cm
\noindent Received September 2003}



\end{document}